\documentclass[prb,showpacs,unsortedaddress,twocolumn,floatfix]{revtex4}

\usepackage{graphicx}
\usepackage{longtable}
\usepackage{graphicx}
\usepackage{array}
\usepackage{amsmath,amsfonts,amssymb}

\begin{document}

  \bibliographystyle{prsty}

  \title{Stark Effect of the Au(111) and Cu(111) Surface States}
  \author{J.\ Kr{\"o}ger}
  \author{L. Limot}
  \author{H.\ Jensen}
  \author{R.\ Berndt}
  \affiliation{Institut f\"ur Experimentelle und Angewandte Physik, Christian-Albrechts-Universit\"at zu Kiel, D-24098 Kiel, Germany}
  \author{P.\ Johansson}
  \affiliation{Department of Natural Sciences, University of \"Orebro, S-70182 \"Orebro, Sweden}

  \date{\today}

  \begin{abstract}
    We present a low-temperature scanning tunneling spectroscopy study of the Au(111) and of the Cu(111)
    surface states showing that their binding energy increases when the
    tip is approached towards the surface. This result, supported by a one-dimensional model calculation
    and by a comparison to existing photoemission spectroscopy measurements,
    confirms the existence of a tip-induced Stark effect as previously reported for Ag(111)
    [L.\ Limot {\it et al.}, Phys.\ Rev.\ Lett.\ {\bf 91}, 196801 (2003)], and suggests that this effect is
    a general feature of scanning tunneling spectroscopy.
  \end{abstract}


  \maketitle
\begin{figure}
    \includegraphics[bbllx=120,bblly=87,bburx=467,bbury=527,width=50mm,clip=]{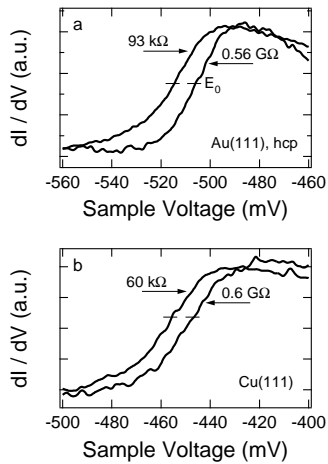}
    \caption[onset_ss]{${\rm d}I/{\rm d}V$ versus sample voltage
    for: (a) Au(111) (hcp region), (b) Cu(111). The tunneling gaps were set at $V=-560\,{\rm mV}$,
    $I=1\,{\rm nA}$ and $V=-600\,{\rm mV}$, $I=1\,{\rm nA}$ for the right-hand-side spectra in (a) and (b),
    respectively, whereas the left-hand-side spectra correspond to gap settings of $V=-560\,{\rm mV}$,
    $I=6\,\mu{\rm A}$ and $V=-600\,{\rm mV}$, $I=10\,\mu{\rm A}$, respectively. The spectra are averages of various
    spectra recorded with different tips and at different locations of the surface.}
    \label{onset_ss}
  \end{figure}
\begin{figure}
    \includegraphics[bbllx=99,bblly=93,bburx=509,bbury=707,width=50mm,clip=]{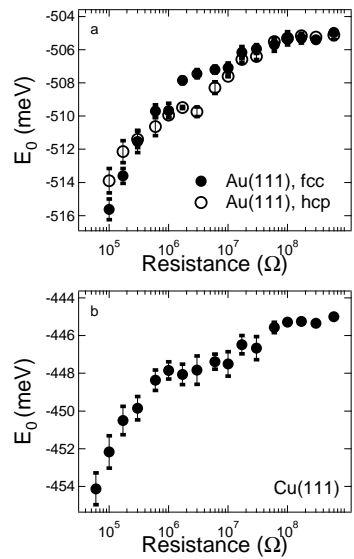}
    \caption[linear]{Surface state binding energy $E_0$ versus the tunnel junction resistance for: (a) Au(111)
    (hcp region: solid circles, fcc region: open circles), (b) Cu(111). Each binding energy is the average value
    of the onset energy extracted from various spectra and the error bar is the corresponding mean deviation.}
    \label{linear}
  \end{figure}
\begin{figure}[]
    \includegraphics[bbllx=88,bblly=92,bburx=500,bbury=680,width=50mm,clip=]{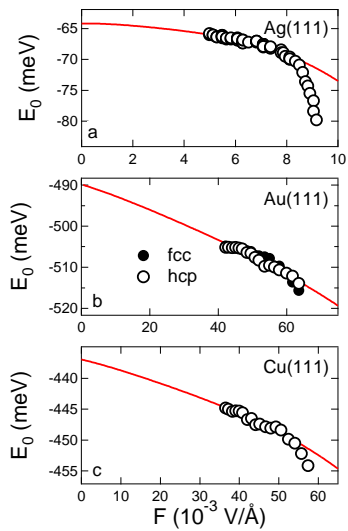}
    \caption[be_dis]{$E_0$ versus the electric field $F$ for (a) Ag(111), (b) Au(111) (hcp region: solid circles,
    fcc region: open circles), and (c) Cu(111), along with the model calculation of the Stark shifts (solid curves).}
    \label{be_dis}
  \end{figure}

  The presence of the electric field between the tip of a scanning tunneling microscope (STM) and the surface is
  known to affect the surface band structure. For instance, when semiconductors are probed with STM, the electric
  field may cause a band bending of the surface electronic structure.\cite{ell93} Contrary to semiconductors, the
  electric field is efficiently screened in metals by the conduction electrons. Nevertheless, the wave function of
  surface state electrons is evanescent into the vacuum, and thus it is prone to be affected by the electric field.
  Indeed, scanning tunneling spectroscopy (STS) studies of field emission resonance or image states by Becker
  {\it et al.}\ and by Binnig {\it et al.}\ showed that these states are affected by a Stark effect.\cite{bec85,bin85}
  However, the field emission states can only be probed by STM at large tip-sample voltages that exceed the tip and
  sample work functions. The question arises whether the Stark effect is still sizeable enough to affect STS spectra
  even in the tunneling regime where STS experiments are usually performed, $i.\, e.$, at voltages considerably lower
  than the tip and the sample work functions. To date, there is a lack of experimental and theoretical data concerning
  this issue.
  \- In this report, we address this topic by a low-temperature STS study of the Au(111) and of the Cu(111) Shockley
  surface states, which represent reference systems for STS. The surface state electrons are an experimental
  realization of a quasi-two-dimensional electron gas spatially localized in the topmost surface layers of a solid,
  their parabolic energy band lying within surface-projected band gaps of the bulk electronic structure. For this
  reason, the investigation of surface states especially by means of angle-resolved photoelectron spectroscopy
  (ARPES),\cite{hei77,kev83,gie85,man88,car90,fra90,rei01} has a long history. Owing to the atomic resolution in
  space and the better-than-meV resolution in energy, STM and STS can map the properties of the surface state
  electrons by imaging their Friedel oscillations arising from the scattering by adsorption atoms and by step
  edges,\cite{dav91,has93,cro93,bur99} and by STS measurements performed over bare surface terraces and in
  artificially engineered nanocavities.\cite{kli00,man00,kli00b,bra02} Overall, there is a good agreement between
  experiments and theory.\cite{ech03}
  \- By varying the electric field of the microscope through the resistance ($R$) of the junction, from
  $1\,{\rm G}\Omega$ down to $60\,{\rm k}\Omega$, we demonstrate that the STS spectra of the Au(111) and of the
  Cu(111) surface states undergo a downward shift in the tunneling regime. We attribute this observation to a Stark
  effect. This result, supported by a one-dimensional model calculation and by a comparison to existing ARPES
  measurements where the tip is absent, confirms the previous observation of a Stark effect in STS of the Ag(111)
  surface state,\cite{lim03} and suggests that this effect is a general feature of STS.
  \- The measurements were performed in a cryogenic STM working at $10\,{\rm K}$ in ultrahigh vacuum (base pressure
  below $10^{-8}\,{\rm Pa}$). We emphasize that the apparatus employed is not the one used previously to investigate
  the Stark effect of the Ag(111) surface state, $i.\, e.$, the present data was acquired using a different microscope
  and different electronics. The sample surfaces were prepared by repeated cycles of argon ion bombardment with
  subsequent annealing. Tunneling spectroscopy of the differential conductance (${\rm d}I/{\rm d}V$) versus the
  sample voltage was performed by opening the feedback loop in the center of areas of
  $300\,{\rm\AA}\times 300\,{\rm\AA}$ where no scattering centers were visible in the topographic images. For the
  Au(111) surface, which exhibits a herringbone reconstruction,
  \cite{per74,mel78,hov81,tak83,har85,wol89,hua90,bar90,cha91} we performed spectroscopic measurements on
  hexagonal-close packed (hcp) and face-centered cubic (fcc) regions of the reconstruction. The differential
  conductance was measured by employing a modulation technique where a sinusoidal voltage (with root-mean-square
  amplitude of $1 - 2\,{\rm mV}$ and a frequency of $\approx 9000\,{\rm Hz}$) is superimposed on the tunneling voltage
  and the current response is measured by a lock-in amplifier. In order to ensure data quality over the range of
  tunneling gap resistances investigated, images of the surface and high-resistance spectra
  ($R\approx 600\,{\rm M}\Omega$) were systematically recorded prior to and after the acquisition of each spectrum,
  following the procedure introduced in Ref.\ \onlinecite{lim03}. Since no change was discernible in these images or
  in these spectra, we conclude that neither a permanent tip modification nor a tip-induced damage of the surface
  occurred when acquiring spectra in the $60\,{\rm k}\Omega$ - $1\,{\rm G}\Omega$ range. We adhered to the procedure
  discussed {\it in extenso} in Ref.\ \onlinecite{lim03} to ensure that artifacts likely to pollute the shift of the
  spectra are negigible (below $0.5\,{\rm meV}$).
  \- Figure\ \ref{onset_ss} presents spectra of the surface states of Au(111)
  (Fig.\ \ref{onset_ss}a) and of Cu(111) (Fig.\ \ref{onset_ss}b) for the lowest and highest
  tunneling resistances where STS was performed. The step-like onset of the differential conductance
  corresponds to the low band edge, or binding energy, $E_{0}$ of the surface state electrons.
  To determine $E_0$, we employ the geometrical analysis of Ref.~\onlinecite{kli00}.
  For the high resistance spectrum, we extract for Au(111) a binding energy of $E_{0}=-505$ meV ($R=560\,{\rm M}\Omega$)
  and for Cu(111) a binding energy of $E_{0}=-445$ meV ($R=600\,{\rm M}\Omega$), in agreement with prior STS measurements.
  \cite{kli00} A sizeable shift of $E_{0}$ for both surfaces is clearly visible in Fig.\ \ref{onset_ss}.
  Upon decreasing $R$ to $93\,{\rm k}\Omega$, the binding energy of the Au(111) surface
  state shifts downward by $\approx 10\,{\rm meV}$, independently on whether the spectrum
  is acquired over a hcp or a fcc region of the herringbone reconstruction (see Fig.\ \ref{linear}).
  Similar to Au(111), the surface state onset of Cu(111) shifts downward by $\approx 9\,{\rm meV}$
  when $R$ is decreased to $60\,{\rm k}\Omega$. No appreciable broadening of the onset is observed.
  \- Figure \ref{linear} is the quantitative evaluation of the band edge shift of all the recorded spectra
  (Fig.\ \ref{linear}a presents the data for Au(111) (the binding energy values for the fcc (hcp) region
  are depicted as full (open) circles); the corresponding data for Cu(111) are presented in Fig.\ \ref{linear}b) .
  As shown,
  the binding energy $E_0$ of the two surface states undergoes a gradual downward shift upon decreasing the resistance.
  For Au(111), the shifts recorded over the hcp and the fcc regions of Au(111) do not exhibit major differences,
  thus the contribution of the Au(111) herringbone reconstruction to the observed shift is negligible.
  Overall, the shift corresponds for both surfaces to a $\approx 2\%$
  variation of $E_{0}$ for $R$ ranging from $100\,{\rm k}\Omega$ to $1\,{\rm G}\Omega$, lower than the $\approx 6\%$
  variation observed for the surface state of Ag(111) over the same range of tunneling resistances.
  \- Finally, in order to express the dependency of $E_{0}$ on the tip displacement and to
  link the data to our model calculation of the shift (presented below), we measured the evolution of $R$
  with the tip-surface separation over a distance of $\approx 6\,{\rm\AA}$ which covers the range of resistances
  of interest. For both surfaces we observe the usual
  tunneling behavior $R\propto\exp(1.025\sqrt{\phi}/d)$, where $d$ is the tip-surface distance, with apparent
  barrier heights of $\phi=(5.2\pm 0.2)\,{\rm eV}$ for Au(111) and of $\phi=(4.7\pm 0.2)\,{\rm eV}$ for Cu(111).
  However, contrary to Au(111), the resistance-versus-displacement curve of Cu(111) exhibits a deviation from
  tunneling behavior for $R\leq 200\,{\rm k}\Omega$, indicating that at these resistances the junction is no longer in a
  tunneling regime; a similar behavior was also observed for Ag(111) when $R\leq 100\,{\rm k}\Omega$.\cite{lim03}
  \- Summarizing the experimental findings, our STS data underline the existence of a downward shift of the Au(111)
  and Cu(111) surface state binding energy upon decreasing the tunneling resistance. Since the resistance
  depends on the the tip-surface
  distance $d$, by varying $R$ we also change the electric field between the tip and the surface ($\propto 1/d$).
  As for Ag(111), the observed increase of the surface state binding energies is therefore a Stark shift
  essentially produced by the
  electric field of the tunnel junction acting on the energy levels of the surface state of Au(111) and of Cu(111).
  \- To emphasize this, we present in Fig.\ \ref{be_dis} the binding energies of the Au(111) (Fig.\ \ref{be_dis}b) and
  of the Cu(111) (Fig.\ \ref{be_dis}c) surface states, along with the binding energies of the Ag(111)
  surface state of Ref.\ \onlinecite{lim03} (Fig.\ \ref{be_dis}a), versus the
  approximate amplitude of the electric field $F=V_{0}/d$,\cite{offset} where we set $eV_{0}$ to the binding energies
  determined by ARPES (see Tab.\ \ref{pes_sts}). The solid lines in Fig.\ \ref{be_dis} are the calculated shifts
  predicted by the one-dimensional model employed in Ref.\ \onlinecite{lim03}. This model is
  based on the surface state potential proposed by Chulkov {\it et al.},\cite{chu99} but is modified to account for
  the presence of the tip by adding to the potential the linear contribution of the voltage between the tip and the
  surface, as well as the difference between the work functions of the tip ($\phi_{\rm t}$) and the surface
  ($\phi_{\rm s}$) to include the contact potential. Furthermore, the shape of the image potential
  is modified to account for multiple images in the tip and the surface.
  \- The calculation reproduces the data over a range of fields, but not at the highest fields investigated where,
  as discussed below, an expected discrepancy occurs with the data. The best fit for Au(111) is achieved with
  $\phi_{\rm t}=5.00\,{\rm eV}$ and $\phi_{\rm s}=5.55\,{\rm eV}$, whereas for Cu(111) and for Ag(111) a work function
  of, respectively, $4.94\,{\rm eV}$ and $4.56\,{\rm eV}$ is sufficient for both tip and surface (all the values of
  $\phi_{\rm s}$ are fixed to those of Ref.\ \onlinecite{chu99}).
  From an experimental point of view, we assume $\phi_{\rm t}\leq\phi_{\rm s}$,
  because the tip is prepared by controlled tip-surface contacts. The tip apex is therefore covered by
  substrate material with a high step density. A lowering of the work function of a gold-coated
  tip by $\approx 0.5\,{\rm eV}$ with respect to the Au(111) work function is consistent with what is observed
  on vicinal gold surfaces with a high step density.\cite{kra77} On the other hand, since for copper
  the Smoluchowski effect is ten times weaker compared to gold, the work function of copper vicinal surfaces
  is lowered typically by $\approx 0.05\,{\rm eV}$ with respect to Cu(111).\cite{rot02} We therefore expect the
  copper-coated tip to yield $\phi_{\rm t}\approx\phi_{\rm s}$, and for Cu(111) the contribution of the contact
  potential to the Stark shift can be neglected. From our model calculation, this is also likely for Ag(111).
  \begin{table}
    \caption{Surface-state binding energies (meV) for the (111) surfaces of Ag, Au, and Cu obtained
    by ARPES and by STS.}
    \begin{ruledtabular}
      \begin{tabular}{lccc}
                                          & Ag(111)               & Au(111)             & Cu(111)             \\
        ARPES$^{\rm a}$                   & $-63\pm 1$            & $-487\pm 1$         & $-435\pm 1$         \\
        STS ($R\to\infty$)                & $-64\pm 1$            & $-490\pm 2$         & $-437\pm 1$         \\
        STS ($R=500\,{\rm M}\Omega$)      & $-66\pm 1$            & $-505\pm 1$         & $-445\pm 1$         \\
      \end{tabular}
    \end{ruledtabular}
    \begin{flushleft}
      $^{\rm a}$ F.\ Reinert {\it et al.}, Phys.\ Rev.\ B {\bf 63}, 115415 (2001). \\
    \end{flushleft}
    \label{pes_sts}
  \end{table}
  \- Focusing first on the low-field behavior of the Stark shift of Fig.\ \ref{be_dis}, it can be seen that when
  $F$ is decreased, $i.\, e.$, when the tip is retracted, the Stark effect progressively disappears and $E_{0}$
  reaches its non-perturbed value. The extrapolation of the
  model calculations to $R\to\infty$ ($F=0$) yields the values of the Stark-free binding energies, which
  agree remarkably well with recent ARPES experiments where no electric field is
  present (Tab.\ \ref{pes_sts}). From our STS data, we conclude that even at the lowest feasible tunneling
  currents which are accessible in experiments, the Stark effect contribution cannot be eliminated.
  As underlined in Tab.\ \ref{pes_sts}, because of the tip-induced Stark effect, the binding
  energies of the noble metal surface states  in usual STS experiments are shifted by a few percent compared to
  their non-perturbed values. More generally, we expect larger Stark shifts for states with wave functions that
  penetrate further into vacuum than these Shockley surface states.
  \- We now turn to the high-field Stark effect. In Figs.\ \ref{be_dis}a and \ref{be_dis}c a deviation
  of the calculated curve from the experimental data is observed starting from $\approx 8\times 10^{-3}\,{\rm V}/{\rm\AA}$ in the
  case of Ag(111) and from $\approx 55\times 10^{-3}\,{\rm V}/{\rm\AA}$ in the case of Cu(111). Increasing the electric field above
  these values leads to an accelerated downward shift of the measured binding energy as compared to theory.
  We attribute this effect to the collapse of the constant tip-surface geometry, which is not accounted for in the
  calculations. It is well known, that at small tip-surface distances, $i.\, e.$, when $R\alt 100\,{\rm k}\Omega$,
  the STM junction is no longer in a tunneling regime, rather in a contact regime where the tip and the surface
  interact.\cite{bra95}
  This leads to local modifications of the tip and of the surface geometries, which must enhance the electric
  field of the STM junction. This conclusion agrees with the fact that the accelerated downward shift for Ag(111) and
  for Cu(111) occurs concomitantly with the deviation from the exponential tunneling behavior seen in
  the resistance-versus-displacement curves. We also note that at these resistances,
  the tip-surface deformations must be reversible since
  the images acquired after the low-resistance spectra did not show any irreversible
  surface modification. The interpretation of an enhanced Stark effect for Cu(111) and for Ag(111)
  in terms of a tip-surface geometry modification, is also consistent
  with the data of Au(111). In fact, for Au(111) there is no enhanced Stark effect (the calculated and the
  experimental shifts match) and an exponential behavior is seen in the resistance-versus-displacement curve
  of Au(111) down to the lowest resistances probed ($R\ge 90\,{\rm k}\Omega$).
  \- In conclusion, we performed low-temperature STS on the Au(111) and on the Cu(111) surface states for tunneling
  resistances ranging from $1\,{\rm G}\Omega$ down to $60\,{\rm k}\Omega$. We observe a downward shift
  of the surface state binding energy upon decreasing the tip-surface distance (the measurements on the hcp and fcc sites
  of the Au(111) herringbone reconstruction yield a similar shift). This energy shift is attributed to a Stark
  effect mainly originating from the tunneling voltage between the tip and the surface,
  as was previously reported for the Ag(111) surface state.\cite{lim03} As for Ag(111), Cu(111) exhibits an enhanced
  Stark effect at low resistances, which we associate to the breakdown of the tunneling regime of the STM.
  The interpretation of the shift in terms of a Stark effect is supported by a one-dimensional model calculation,
  from which we extract Stark-free binding energies close to those determined in the ARPES experiments of
  Ref.\ \onlinecite{rei01}. The presence of a Stark effect in reference systems such as the noble metal surface
  states, even at usual tunneling parameters, strongly suggests that this effect is quite common for STM and for STS.
  \- J.\ K., L.\ L., H.\ J., and R.\ B.\ acknowledge financial support by the Deutsche Forschungsgemeinschaft, and
  P.\ J.\ by the Swedish Research Council.


\begin{thebibliography}{nnnyys}
\bibitem{offset} Our model calculation yields $d\approx 12\,{\rm\AA}$ for
                     $R=1\,{\rm G}\Omega$. We employ this reference position to
                     estimate the absolute tip-surface distance. A variation of $\pm 1\,{\rm\AA}$ has only a
                     negligible impact on the analysis of the data, accounted for in the error bars of
                     Tab.\ \ref{pes_sts}.
\end{thebibliography}
\end{document}